\documentclass[12pt]{article}
\usepackage{amsmath}
\usepackage{amsfonts,amssymb}
\usepackage{latexsym}
\usepackage{graphics}
\def\d{\mbox{\rm d}}

\def\half{\mbox{$\frac{1}{2}$}}
\begin{document}
\title{Gauge Variant Symmetries for the Schr\"odinger Equation}
\author{M.C. Nucci and P.G.L. Leach\footnote{permanent address: School of Mathematical Sciences, Westville Campus, University of
KwaZulu-Natal, Durban 4000, Republic of South Africa}}
\date{Dipartimento di Matematica e Informatica\\ Universit\`a di
Perugia, 06123  Perugia, Italy}

\maketitle
\begin{abstract}
The last multiplier of Jacobi provides a route for the
determination of families of Lagrangians for a given system.  We
show that the members of a family are equivalent in that they
differ by a total time derivative.  We derive the Schr\"odinger
equation for a one-degree-of-freedom system with a constant
multiplier.  In the sequel we consider the particular example of
the
 simple harmonic oscillator. In the case of the general equation for the
simple harmonic oscillator which contains an arbitrary function we
show that all Schr\"odinger equations possess the same number of
Lie point symmetries with the same algebra.  From the symmetries
we construct the solutions of the Schr\"odinger equation and find
that they differ only by a phase determined by the
gauge.\end{abstract} {\bf PACS numbers}: 02.20.-a; 02.30.-f;
03.65.Fd; 03.65.Ge\\ {\bf Keywords}: Jacobi's last multiplier, Lie
symmetry, creation and annihilation operators.

\strut\hfill

\section{Introduction}
\label{intro} Jacobi's Last Multiplier is a solution of the linear
partial differential
 equation \cite {Jacobi 86 a, Whittaker 44 a},
\begin {equation}
\frac {\partial M} {\partial t} +\sum_{i = 1} ^N \frac {\partial
(Ma_i)} {\partial x_i} = 0, \label{1.0}
\end {equation}
where $\partial_t+\sum_{i = 1} ^N a_i\partial_{x_i} $ is the
vector field of the set of first-order ordinary differential
equations for the $N $ dependent variables $x_i $.    The
relationship between the Jacobi Last Multiplier and the
Lagrangian, {\it videlicet} \cite {Jacobi 86 a,Whittaker 44 a}
\begin {equation}
\frac {\partial ^ 2L} {\partial\dot{x} ^ 2} = M \label {1.1}
\end {equation}
for a one-degree-of-freedom system, is perhaps not widely known
although it is certainly not unknown as can be seen from the
bibliography in \cite {Nucci 05 a}.  If two multipliers, $M_1 $
and $M_2 $, are known, their ratio is a first integral.  In the
case of a conservative system with the standard energy integral
\begin {equation}
E = \half\dot {x} ^ 2+ V (x) \label {1.2}
\end {equation}
and Lagrangian
\begin {equation}
L = \half\dot {x} ^ 2- V (x) \label {1.3}
\end {equation}
it is evident from (\ref {1.1}) that one multiplier is a constant
-- taken to be 1 without loss of generality -- and so all
multipliers are first integrals.  This combined with (\ref {1.1})
is a simple recipe for the generation of a Lagrangian.  One has
\begin {equation}
\frac {\partial ^ 2L} {\partial\dot{x} ^ 2} =
1\quad\Longrightarrow\quad L = \half\dot {x} ^ 2+ \dot {x}f_1
(t,x) + f_2 (t,x), \label {1.4}
\end {equation}
where $f_1 $ and $f_2 $ are arbitrary functions of integration.
Naturally different multipliers give rise to different
Lagrangians.  For a study of these in the classical context with
particular reference to their inequivalence and Noether symmetries
see \cite {Nucci 07 a}.  Lagrange's equation of motion for (\ref
{1.4}) is
\begin {equation}
\ddot {x} + \frac {\partial f_1} {\partial t} - \frac {\partial
f_2} {\partial x} = 0 \label {1.5}
\end {equation}
whereas that for (\ref {1.3}) is
\begin {equation} \ddot {x} + V'(x) = 0 \label {1.6}
\end {equation}
so that the requirement that the two Newtonian equations be the
same is
\begin {equation}
\frac {\partial f_1} {\partial t} - \frac {\partial f_2} {\partial
x} = V' (x ) \label {1.7}
\end {equation}
which is a constraint on the doubly infinite family of functions,
$f_1 $ and $f_2 $.  This constraint may be expressed through
setting
\begin {equation}
f_1 = \frac {\partial g} {\partial x},\quad f_2 = \frac {\partial
g} {\partial t} - V (x), \label {1.8}
\end {equation}
where $g (t,x) $ is an arbitrary function of its arguments.
Consequently the Lagrangian, (\ref {1.4}), becomes
\begin {eqnarray}
L & = & \half\dot {x} ^ 2- V (x) + \dot {x}\displaystyle {\frac {\partial g} {\partial x} +\frac {\partial g} {\partial t}} \nonumber \\
& = & \half\dot {x} ^ 2- V (x) + \dot {g}, \label {1.9}
\end {eqnarray}
{\it ie}, the functions $f_1 $ and $f_2 $ are a consequence of the
arbitrariness of a Lagrangian with respect to a total time
derivative, the gauge function.

\strut\hfill

Although we confine our attention to equations of motion such as
(\ref {1.6}) with a constant multiplier, the discussion above is
quite general.  By way of example consider the Hamiltonian \cite
{Goldstein 80 a} [ex 18, p 433]
\begin {equation}
H = \half\left (\frac {1} {q ^ 2} +p ^ 2q ^ 4\right) \label {1.10}
\end {equation}
with Lagrangian and equation of motion
\begin {equation}
L = \half\left (\frac {\dot {q} ^ 2} {q ^ 4} -\frac {1} {q ^
2}\right)\quad\quad q\ddot {q} - 2\dot {q} ^ 2-q ^ 2 = 0.  \label
{1.11}
\end {equation}
From the Lagrangian and (\ref {1.1}) it is evident that a
multiplier is
\begin {equation}
M = \frac {1} {q ^ 4} \label {1.12}
\end {equation}
and so the general Lagrangian is
\begin {equation}
L = \half\frac {\dot {q} ^ 2} {q ^ 4} +\dot {q}f_1 (t,q) +f_2
(t,q). \label {1.13}
\end {equation}
The precise equation of motion in (\ref {1.11}) is obtained if
$f_1 $ and $f_2 $ are constrained according to
\begin {equation}
\frac {\partial f_1} {\partial t} -\frac {\partial f_2} {\partial
q} = -\frac {1} {q ^ 3}. \label {1.14}
\end {equation}
If we write
\begin {equation}
f_1 = \frac {\partial g} {\partial x}\quad\mbox {\rm and}\quad f_2
= -\frac {\partial g} {\partial t} -\frac {1} {2q ^ 2}, \label
{1.15}
\end {equation}
we clearly see that the imposition of the constraint (\ref {1.14})
again introduces an arbitrary gauge.  We note that the
Hamiltonian, (\ref {1.10}), is connected by a canonical
transformation, specifically $Q= -q^ {- 1},\,P= q^2p$ yields $K= \mbox {$\frac {1} {2}$}\left (P^2+Q^2\right) $, to the standard Hamiltonian of the simple harmonic
oscillator.

\strut\hfill

 Naturally we do not have to confine our attention to linear problems.  For example the simple pendulum with Newtonian equation of motion
\begin {equation}
\ddot {q} +\omega ^ 2\sin q = 0 \label {1.16}
\end {equation}
has a Lagrangian of the form given in (\ref {1.4}), but now the
constraint on the functions $f_1 $ and $f_2 $ is
\begin {equation}
\frac {\partial f_1} {\partial t} -\frac {\partial f_2} {\partial
q} = \omega ^ 2\sin q \label {1.17}
\end {equation}
which can be satisfied if we introduce a gauge function, $g (t,q)
$, such that
\begin {equation}
f_1 = \frac {\partial g} {\partial q}\quad\mbox {\rm and}\quad f_2
= \frac {\partial g} {\partial t} -\omega ^ 2\left (1-\cos
q\right). \label {1.18}
\end {equation}

\strut\hfill

In this paper we wish to explore the implications of generality of
the Lagrangian, (\ref {1.4}), in the context of the corresponding
Schr\"odinger Equation.  To make the work quite explicit we use
the simple harmonic oscillator as a vehicle.  There is a simple
reason for this choice.  The simple harmonic oscillator is richly
endowed with symmetry which are the eight Lie point symmetries of
its Newtonian equation of motion, the five Noether point
symmetries of its Action Integral and the five plus one plus
infinity Lie point symmetries of its Schr\"odinger Equation.  The
last provide an algorithmic route to the determination of the
wave-functions \cite {Lemmer 99 a,Andriopoulos 05 a,Leach 06
a,Andriopoulos 06 a}. What we find here is that the Schr\"odinger
Equation of the Hamiltonian corresponding to the Lagrangian with
an arbitrary gauge function has the same number of Lie point
symmetries as the standard Schr\"odinger Equation for the simple
harmonic oscillator.  However, the unknown functions, $f_1 $ and
$f_2 $ are present in the symmetries. Nevertheless the
determination of the wave-functions using these Lie point
symmetries proceeds without hindrance.

\strut\hfill

\section{Schr\"odinger Equation}
\label{sec:1}

\strut\hfill

The canonical momentum for (\ref {1.4}) is
\begin {equation}
p = \frac {\partial L} {\partial\dot {x}} = \dot {x} + f_1 \label
{2.1}
\end {equation}
so that
\begin {eqnarray}
H & = & p\dot {x} - L \nonumber\\
& = & \half p ^ 2- pf_1+ \half f_1 ^ 2- f_2 \label {2.2}
\end {eqnarray}
is the Hamiltonian.  Whether one uses the Weyl quantisation
formula or the symmetrisation of $pf_1 $ make no difference to the
form of the Schr\"odinger Equation corresponding to (\ref {2.2})
which is
\begin {equation}
2i\frac {\partial u} {\partial t} = -\frac {\partial ^ 2u}
{\partial x ^ 2} + 2 if_1\frac {\partial u} {\partial x} +\left
(f_1 ^ 2-2f_2+i\frac {\partial f_1} {\partial x}\right)u. \label
{2.3}
\end {equation}

\strut\hfill

The Schr\"odinger Equation, (\ref {2.3}), is quite general.  We
now introduce the simple harmonic oscillator with Newtonian
equation of motion $\ddot {x} +k ^ 2x = 0 $ so that the constraint
(\ref {1.7}) is
\begin {equation}
\frac {\partial f_1} {\partial t} - \frac {\partial f_2} {\partial
x} = k ^ 2x.  \label {2.4}
\end {equation}
One notes that neither of the derivatives in the constraint
appears in the Schr\"odinger Equation, (\ref {2.3}).  As we
observed above, the reason for using the simple harmonic
oscillator as the explicit example is simply due to its generous
supply of point symmetries.

\strut\hfill

The Lie point symmetries of (\ref {2.3}) subject to the constraint
(\ref {2.4}) are\footnote {Thanks to Nucci's interactive package
for the computation of Lie symmetries \cite {Nucci 90 a,Nucci 96
a}.}
\begin{eqnarray}
\Gamma_1 &=& \cos(k t)\,\partial_x+ \left[\cos(k t) f_1 - \sin(k t)  k  x\right] iu\,\partial_u   \nonumber\\
\Gamma_2 &=& - \sin(k t)\,\partial_x-\left[ \sin(k t) f_1+\cos(k t) k x\right ] i u\, \partial_u  \nonumber\\
\Gamma_3 &=& \partial_t+\left( f_2 +\frac{1}{2} \, k^2  x^2\right) iu\,\partial_u \nonumber\\
\Gamma_4 &=& \cos(2 k t)\,\partial_t -\sin(2 k t) k x\, \partial_x\nonumber\\
  &&+\left[i \cos(2 k t) \left( f_2 -\frac{1}{2}\, k^2 x^2\right)   - k\sin(2 k t)\left(  i   x f_1 -\frac{1}{2}\right)\right]u\partial_u   \nonumber\\
\Gamma_5 &=& - \sin(2 k t)\partial_t -\cos(2 k t) k x \partial_x\nonumber\\
  &&-\left[i\sin(2 k t) \left(f_2 -\frac{1}{2}\, k^2 x^2\right)   + k \cos(2 k t) \left( i  x f_1-\frac{1}{2}\right)  \right]u\partial_u   \nonumber\\
\Gamma_6 &=& u \,\partial_u \nonumber \\
\Gamma_7 & = & s (t,x)\partial_u, \label {2.5}
\end{eqnarray}
where $s (t,x) $ is a solution of (\ref {2.3}), which is a
representation of the well-known algebra, $\{sl (2,R)\oplus_s
W\}\oplus_s\infty A_1 $, of the Schr\"odinger Equation for the
one-dimensional linear oscillator and related systems.  The
presence of the functions $f_1 $ and $f_2 $ subject to the
constraint (\ref {2.4}) does not affect the number of Lie point
symmetries of (\ref {2.3}) vis-\`a-vis the number for the
Schr\"odinger Equation for the simple harmonic oscillator\footnote
{Although we make no attempt to prove it, one can easily believe
that the result is independent of the particular potential.  It is
just that the Schr\"odinger Equation with a general potential is
rather lacking in Lie point symmetries apart from $\Gamma_6 $ and
$\Gamma_7 $.  We emphasise, however, that it is not invariably the
case that the Schr\"odinger Equation constructed from a given
Hamiltonian has the same number of Lie point symmetries as the
corresponding Lagrangian has Noether point symmetries plus
$\Gamma_6 $ and $\Gamma_7 $ which are a consequence of the
linearity of the equation.}.  The symmetries listed in (\ref
{2.5}) reduce to those for the standard Schr\"odinger Equation of
the simple harmonic oscillator if one sets, say, $f_1 = 0 $ and
$f_2 = -\half k ^ 2x ^ 2 $, which is an obvious solution of (\ref
{2.4}).

\strut\hfill

\section {Creation and Annihilation Operators}
\label{sec:2}

For the purposes of quantum mechanics one usually writes
symmetries of the structure of $\Gamma_1 $ and $\Gamma_2 $ and
$\Gamma_4 $ and $\Gamma_5 $ in terms of an exponential rather than
trigonometric functions.  Thus we write
\begin {eqnarray}
\!\!\!\!\!\!\!\!\!\!\Gamma_{1\pm} & = & \exp [\pm
kit]\{\partial_x+i (f_1\pm i kx)u\partial_u\} \label {2.6}
\\\!\!\!\!\!\!\!\!\!\!
\Gamma_{4\pm} & = & \exp [\pm 2 kit]\left\{\partial_t \pm
kix\partial_x+i\left [\left (f_2-\half k ^ 2x ^ 2\right) \pm
k\left (2if_1-\half \right)\right]u\partial_u\right\}\!. \label
{2.7}
\end {eqnarray}
The normal route to the solution of the Schr\"odinger Equation,
(\ref {2.3}), is to use the symmetries (\ref {2.6})\footnote
{These are, as equally $\Gamma_1 $ and $\Gamma_2 $, often termed
`solution symmetries' since they correspond to the solutions of
the corresponding Newtonian equation of motion.  For the
Schr\"odinger Equation the symmetries $\Gamma_7 $ are {\it the}
solution symmetries, but they do not play the same role as
$\Gamma_1 $ and $\Gamma_2 $ or $\Gamma_{1\pm} $ which is the
specification of the potential.} which are the time-dependent
progenitors of the well-known creation and annihilation operators
of Dirac in the case that $f_1 $ and $f_2 $ are restricted as
above.

\strut\hfill

To solve the Schr\"odinger Equation, (\ref {2.3}), using Lie's
method we reduce (\ref {2.3}) to an ordinary differential equation
by using the invariants of the symmetries as a source of the
variables.  We must also be cognisant of the need for the solution
of (\ref {2.3}) to satisfy the boundary conditions at $\pm\infty
$.  With this requirement in mind we take $\Gamma_{1+} $.  The
associated Lagrange's system is
\begin {equation}
\frac {\d t} {0} = \frac {\d x} {1} = \frac {\d u} {i\left
(f_1+kix\right)u} \label {2.8}
\end {equation}
which gives the characteristics $t $ and $u\exp\left [\half kx ^
2-ig (t,x)\right] $, where we have made use of the first of (\ref
{1.8}) and the fact that $t $ is a characteristic.  To find the
solution corresponding to $\Gamma_{1+} $ we write
\begin {equation}
u (t,x) = h (t)\exp\left [-\half kx ^ 2+ig (t,x)\right], \label
{2.9}
\end {equation}
where $h (t) $ is to be determined, and substitute it into (\ref
{2.3}) which simplifies to
$$
i\dot{h} = \half kh
$$
so that
$$
h (t) = \exp\left [-\half kit\right]
$$
and
\begin {equation}
u (t,x) = \exp\left [-\half kit-\half kx ^ 2+ig (t,x)\right]
\label {2.10}
\end {equation}
up to a normalisation constant which we ignore.  With $g = 0 $ we
recognise the ground-state solution for the time-dependent
Schr\"odinger Equation of the simple harmonic oscillator.

\strut\hfill

Evidently we use $\Gamma_{1-} $ as a time-dependent `creation
operator'.  If we write the left hand side of (\ref {2.10}) as
$u_0 $, we can have a solution symmetry of the form
\begin {equation}
\Gamma_{70} = u_0 (t,x)\partial_u, \label {2.11}
\end {equation}
where the subscript, 7j, means that we are using the symmetry
$\Gamma_7 $ with the specific solution, $u_j (t,x) $. We use the
closure of the Lie algebra under the operation of taking the Lie
Bracket to obtain further solutions.  Thus
\begin {equation}
\left [\Gamma_{1-},\,\Gamma_{70}\right]_{LB}  =  \left\{-kx
+i\displaystyle {\frac {\partial g} {\partial x}} - (if_1
+kx)\right\}\exp\left [-\mbox {$\frac {3} {2} $} kit -\half kx ^
2+ig\right]\partial_u \label {2.12}
\end {equation}
so that we have
\begin {equation}
u_1 (t,x) = - 2kx \exp\left [-\mbox {$\frac {3} {2} $} kit -\half
kx ^ 2+ig\right]. \label {2.13}
\end {equation}
Likewise $\left [\Gamma_{1-},\,\Gamma_{71}\right]_{LB}  $ gives
\begin {equation}
u_2 (t,x) = \left (4k ^ 2x ^ 2- 2k \right)\exp\left [-\mbox
{$\frac {3} {2} $} kit -\half kx ^ 2+ig\right]. \label {2.14}
\end {equation}
On the other hand the Lie Bracket of $\Gamma_{4-} $ with
$\Gamma_{70} $ gives
\begin {equation}
\tilde {u}_2 (t,x) = \left (-ik + 2ik ^ 2x ^ 2\right)\exp\left
[-\mbox {$\frac {3} {2} $} kit -\half kx ^ 2+ig\right], \label
{2.15}
\end {equation}
which is $u_2 $ up to a constant factor, {\it ie} $\Gamma_{4-} $
acts as a double creation operator.  One is not surprised that
$\Gamma_{4+} $ is a double annihilation operator.  Obviously
$\Gamma_{1+} $, since it was used to derive the ground state, is
the annihilation operator.

\strut\hfill

Finally the Lie Bracket of $i\Gamma_3 $ with $\Gamma_7 $ yields
the energy.  For example with $\Gamma_{72} $ one has
\begin {equation}
\left [i\Gamma_{3},\,\Gamma_{72}\right]_{LB}  =  \mbox {$\frac {5}
{2} $}ku_2\partial_u. \label {2.16}
\end {equation}

\strut\hfill

In general the action of $\Gamma_{1 -} $ on a solution symmetry $\Gamma_{7j} $, which has been generated by the $j $-fold action of $\Gamma_{1 -} $ on $\Gamma_{70} $ by means of the taking of the Lie Bracket, is
\[
\left [\Gamma_{1 -},\,\Gamma_{7j}\right]_{LB} = \Gamma_{7,,j+1}
\]
and the energy eigenvalue is given by the action of $i\Gamma_3 $ as
\begin {eqnarray*}
\left [\Gamma_{3},\,\Gamma_{7j}\right]_{LB} & = & (j+ \mbox {$\frac {1} {2} $})ku_j\partial_u \\
& = & (j+ \mbox {$\frac {1} {2} $})\Gamma_{7,,j}.
\end {eqnarray*}
It should be quite evident that the operators $\Gamma_{1\pm} $ are the sources of the creation and annihilation operators introduced by Dirac for the time-independent Schr\"odinger equation of the simple harmonic oscillator.  Similar operators are found for time-dependent quadratic Hamiltonians and these play roles similar to those played by the operators reported here.

\strut\hfill

\section {Discussion}
\label{sec:3}

If one derives a Jacobi Last Multiplier for a
one-degree-of-freedom system, the connection, (\ref {1.1}),
between the multiplier and the Lagrangians leads to a doubly
infinite family of Lagrangian for the same multiplier and so a
doubly infinite family of Lagrangian equations of motion.
Insistence that the Lagrangians be specific to the given Newtonian
equation of motion still leaves an infinite class of Lagrangians
related by a gauge function.

\strut\hfill

In the case of a constant multiplier one obtains a Lagrangian
quadratic in the velocity and hence an Hamiltonian quadratic in
the momentum.  From the Hamiltonian one can construct a
Schr\"odinger Equation of quite general form.  To enable the
obtaining of precise results we specialised to the simple harmonic
oscillator.  We found that the number of Lie point symmetries of
the Schr\"odinger Equation was unchanged from that of the standard
Schr\"odinger Equation for the simple harmonic oscillator.  The
algebra is the same even with the presence of $f_1 $ and $f_2 $
subject to the single constraint (\ref {2.4}).  Using the
exceptional Lie point symmetries in the combinations,
$\Gamma_{1\pm} $, $\Gamma_{4\pm} $ and $\Gamma_3 $, we were able
to construct solutions for the Schr\"odinger Equation and found
that these Lie point symmetries acted as creation, annihilation
and energy operators.

\strut\hfill

Classically the presence of the gauge function does not affect the
form of the Lagrangian equation of motion.  In the case of the
simple harmonic oscillator we saw that the wave functions and
energy levels are the same as when the usual Schr\"odinger
Equation for the simple harmonic oscillator is used with the
exception of a phase, $\exp [ig (t,x)] $, which otherwise does not
intrude.

\strut\hfill

It is a natural question to enquire of the extension of the considerations in this paper to the Hamiltonians of the form
\begin {equation}
H = \mbox {$\frac {1} {2} $}p ^2 + V (q,t)\quad\Leftrightarrow\quad L =  \mbox {$\frac {1} {2} $}\dot {q} ^2 - V (q,t) \label {4.1}
\end {equation}
for which the Jacobi Last Multiplier is also $1$ as calculated by means of a reversal of (\ref {1.1}).  In the case that $V $ is quadratic in $q $ the results reported here persist {\it mutatis mutandis} since there is no change in the underlying basis of symmetry \cite {Andriopoulos 06 a}.  For a general potential as in (\ref {4.1}) there is a constant multiplier which means that any other multiplier is an integral.  However, the calculation of additional multipliers is either by the solution of (\ref {1.0}) or the method of Lie using the symmetries of the Schr\"odinger equation.  Neither method is fruitful for a general $V (t,q) $.  The relationship between a Jacobi Last Multiplier and a Lagrangian is quite specific.  A Lagrangian obtained from a Jacobi Last Multiplier and consistent with the equations of motion is under no obligation to possess a sufficient number of first integrals for the Theorem of Liouville to apply.  The absence of symmetry in the Euler-Lagrange equation means an absence of Jacobi Last Multipliers apart from the one which generates (\ref {4.1}) (in this case).  Consequently there is an absence of integrals.  The one multiplier gives the Lagrangian.  The absence of others denies integrability.

\strut\hfill

\section{Acknowledgments}
This work was undertaken while PGLL was enjoying the hospitality
of Professor MC Nucci and the facilities of the Dipartimento di
Matematica e Informatica, Universit\`a di Perugia.  The continued
support of the University of KwaZulu-Natal is gratefully
acknowledged.

\strut\hfill


\begin {thebibliography} {99}

\bibitem {Andriopoulos 05 a}
Andriopoulos, K. and Leach, P.G.L.: {Wave-functions for the
time-dependent linear oscillator and Lie point symmetries}. {
Journal of Physics A: Mathematical and General} {\bf 38},
4365-4374 (2005)

\bibitem {Andriopoulos 06 a}
Andriopoulos, K. and Leach, P.G.L.: {Lie point symmetries: An
alternative approach to wave-functions}. {Bulletin of the Greek
Mathematical Society} {\bf 52} 25-34

\bibitem {Goldstein 80 a}
Goldstein, Herbert: {\it Classical Mechanics}.  Reading MA:
Addison-Wesley (2nd ed), 1980

\bibitem {Jacobi 86 a}
Jacobi, C.G.J.: {\it Vorlesungen \"uber Dynamik.  Nebst f\"unf
hinterlassenen Abhandlungen desselben herausgegeben  von A
Clebsch}.  Berlin: Druck und Verlag von Georg Reimer, 1886

\bibitem {Lemmer 99 a}
Lemmer, R.L. and Leach, P.G.L.: {A classical viewpoint on quantum
chaos}. {Arab Journal of Mathematical Sciences} {\bf 5}, 1--17
(1999/1420)

\bibitem{Levy-Leblond 71 a}
L\'evy-Leblond, Jean-Marc: {Conservation laws for gauge-variant
Lagrangians in Classical Mechanics}. {American Journal of Physics}
{\bf 39}, 502--506 (1971)

\bibitem {Nucci 90 a}
Nucci, M.C.: {Interactive REDUCE programs for calculating
classical, nonclassical and Lie-B\"acklund symmetries for
differential equations}.  Preprint: Georgia Institute of
Technology, Math 062090-051 (1990)

\bibitem {Nucci 96 a}
Nucci, M.C.: {Interactive REDUCE programs for calculating Lie
point, nonclassical, Lie-B\"{a}cklund, and approximate symmetries
of differential equations: manual and floppy disk} in {\it CRC
Handbook of Lie Group Analysis of Differential Equations. Vol.
III: New Trends in Theoretical Developments and Computational
Methods},  Ibragimov NH ed. Boca Raton: CRC Press, 415--482 (1996)

\bibitem {Nucci 05 a}
Nucci, M.C.: {Jacobi last multiplier and Lie symmetries:  a novel
application of an old relationship}. {Journal of Nonlinear
Mathematical Physics} {\bf 12}, 284--304 (2005)

\bibitem {Leach 06 a}
Nucci, M.C., Leach, P.G.L. and Andriopoulos, K.: {Lie symmetries,
quantisation and $c $-isochronous nonlinear oscillators}. {
Journal of Mathematical Analysis and Applications} {\bf 319},
357--368 (2007)

\bibitem {Nucci 07 a}
Nucci, M.C. and Leach, P.G.L.: {Lagrangians galore}.
{ArXiv:0706.1008v1 [nlin.SI]} (2007).

\bibitem{Whittaker 44 a}
Whittaker, E.T.: {\it A Treatise on the Analytical Dynamics of
Particles and Rigid Bodies}. New York: Dover, 1944

\end {thebibliography}
\end {document}